\documentclass[aps,prd,twocolumn,floatfix,altaffilletter,superscriptaddress,preprintnumbers,
tightenlines,showpacs,showkeys,nofootinbib,notoccite]{revtex4-1}
\usepackage[utf8]{inputenc}
\usepackage[colorlinks=true,citecolor=blue,linkcolor=blue]{hyperref}
\usepackage[normalem]{ulem}
\usepackage{amsmath,amssymb, mathrsfs,esvect}
\usepackage{epsfig}
\usepackage{graphicx}               
\usepackage{url}
\usepackage{color}
\usepackage{slashed}
\usepackage{multirow}
\usepackage{placeins}
\usepackage[dvipsnames]{xcolor}
\usepackage{epstopdf}
\usepackage{soul}
\usepackage{tikz}
\usepackage[capitalise, english]{cleveref}
\usepackage{siunitx}
\usepackage{xspace}
\usepackage{booktabs}
\usepackage[capitalise]{cleveref}
\usetikzlibrary{trees}
\usetikzlibrary{decorations.pathmorphing}
\usetikzlibrary{decorations.markings}

\newcommand\myshade{80}
\colorlet{mylinkcolor}{ForestGreen}
\colorlet{mycitecolor}{Red}
\colorlet{myurlcolor}{violet}

\hypersetup{
	linkcolor  = mylinkcolor!\myshade!black,
	citecolor  = mycitecolor!\myshade!black,
	urlcolor   = myurlcolor!\myshade!black,
	colorlinks = true
}

\definecolor{jblue}{RGB}{20,50,100}
\definecolor{npurple}{RGB} {153, 51, 204}
\definecolor{wred}{RGB}{217,0,56}
\definecolor{white}{RGB}{255,255,255}

\allowdisplaybreaks

\usepackage{tikz,xcolor,hyperref}

\definecolor{lime}{HTML}{A6CE39}
\DeclareRobustCommand{\orcidicon}{\hspace{-1mm}
	\begin{tikzpicture}
		\draw[lime, fill=lime] (0,0) 
		circle [radius=0.16] 
		node[white] {{\fontfamily{qag}\selectfont \tiny \,ID}};
		\draw[white, fill=white] (-0.0525,0.095) 
		circle [radius=0.007];
	\end{tikzpicture}
	\hspace{-3mm}
}

\foreach \x in {A, ..., Z}{\expandafter\xdef\csname orcid\x\endcsname{\noexpand\href{https://orcid.org/\csname orcidauthor\x\endcsname}
		{\noexpand\orcidicon}}
}

\keywords{}

\def\lvk{LIGO, Virgo and KAGRA\xspace}
\def\gw{GW\xspace}
\def\gwh{GW\xspace}

\def\cws{continuous waves\xspace}
\def\gws{gravitational waves\xspace}

\def\dm{DM\xspace}
\def\bh{BH\xspace}
\def\bhs{BHs\xspace}
\def\dmh{DM\xspace}

\def\pbhs{primordial black holes\xspace}

\def\lvk{LIGO, Virgo and KAGRA\xspace}

\def\ifos{interferometers\xspace}
\def\invkpccubedyr{kpc$^{-3}$yr$^{-1}$\xspace}
\def\tbh{TBH\xspace}

\def\snr{signal-to-noise ratio\xspace}
\def\fgw{f_{\rm GW}}

\newcommand{\bea}{\begin{eqnarray}}
	\newcommand{\eea}{\end{eqnarray}}
\newcommand{\be}{\begin{equation}}
	\newcommand{\ee}{\end{equation}}

\newcommand{\avgVT}{\ensuremath{\left\langle VT \right\rangle}}

\newcommand{\TFFT}{T_\text{FFT}}

\newcommand{\Tobs}{t_\text{obs}}

\newcommand{\msun}{\ensuremath{M_\odot}\xspace}

\newcommand{\rorb}{R_{\text{orb}}}

\begin{document}
	
	\title{Continuous Gravitational Waves: A New Window to Look for \\ Heavy Non-annihilating Dark Matter}
	\author{Sulagna Bhattacharya\orcidA}
	\email{sulagna@theory.tifr.res.in}
	\affiliation{Tata Institute of Fundamental Research, Homi Bhabha Road, Mumbai 400005, India}
	\author{Andrew L. Miller\orcidB}
	\email{andrew.miller@nikhef.nl}
	\affiliation{Nikhef -- National Institute for Subatomic Physics,
		Science Park 105, 1098 XG Amsterdam, The Netherlands}
	\affiliation{Institute for Gravitational and Subatomic Physics (GRASP),
		Utrecht University, Princetonplein 1, 3584 CC Utrecht, The Netherlands}
	\author{Anupam Ray\orcidC}
	\email{anupam.ray@berkeley.edu}
	\affiliation{Department of Physics, University of California Berkeley, Berkeley, California 94720, USA}
	\affiliation{School of Physics and Astronomy, University of Minnesota, Minneapolis, MN 55455, USA}
	
	\date{\today}
	
	\begin{abstract}
		Sun-like stars can transmute into comparable mass black holes by steadily accumulating heavy non-annihilating dark matter particles over the course of their lives. If such stars form in binary systems, they could give rise to quasi-monochromatic, persistent gravitational waves, commonly known as continuous gravitational waves, as they inspiral toward one another. We demonstrate that next-generation space-based detectors, e.g.,  Laser Interferometer Space Antenna (LISA) and Big Bang Observer (BBO), can provide novel constraints on dark matter parameters (dark matter mass and its interaction cross-section with the nucleons) by probing gravitational waves from transmuted Sun-like stars that are in close binaries. Our projected constraints depend on several astrophysical uncertainties, and nevertheless, are competitive with the existing constraints obtained from cosmological measurements as well as terrestrial direct searches, demonstrating a notable science-case for these space-based gravitational wave detectors as probes of particle dark matter.
	\end{abstract}
	\maketitle
	\preprint{TIFR/TH/24-1, N3AS-24-009}
	\section{Introduction}
	\lvk\cite{TheLIGOScientific:2014jea,TheVirgo:2014hva,KAGRA:2020tym} have observed the $\mathcal{O}(100)$ binary black hole (BH) and neutron star mergers \cite{LIGOScientific:2021djp,Venumadhav:2019lyq,Zackay:2019btq,Olsen:2022pin} in their various observing runs over the past few years. However, another class of gravitational waves (GWs), commonly known as, ``continuous gravitational waves'', have so far eluded detection. Continuous waves are characterized as quasi-monochromatic (a sinusoid with a tiny frequency drift over time) as well as quasi-infinite  (a signal whose duration greatly exceeds the observation time), and can be emitted from an array of sources, e.g., asymmetrically rotating isolated neutron stars \cite{Riles:2017evm, Tenorio:2021wmz, Piccinni:2022vsd, Riles:2022wwz,Miller:2023qyw}, millisecond pulsars at the galactic center \cite{KAGRA:2022osp,Miller:2023qph}, annihilating ultralight dark matter (\dmh) clouds around rotating \bhs \cite{DAntonio:2018sff, Isi:2018pzk,Sun:2019mqb,LIGOScientific:2021jlr}, inspiraling planetary-mass \pbhs \cite{Miller:2018rbg,Miller:2020kmv,Miller:2021knj,Guo:2022sdd, Andres-Carcasona:2023zny,Alestas:2024ubs, Miller:2024fpo}, and even ultralight particle \dm that directly couples to the \ifos \cite{Pierce:2018xmy,Guo:2019ker,Miller:2020vsl,Michimura:2020vxn, LIGOScientific:2021odm,Vermeulen:2021epa,Miller:2022wxu,Miller:2023kkd,Manita:2023mnc}.\\
	In this work, we point out the exciting possibility that continuous gravitational wave measurements can also probe strongly-interacting heavy non-annihilating \dm. Such a DM model is hard to probe in the terrestrial detectors because of their tiny fluxes (DM flux in the terrestrial detector scales inversely with its mass) and remains to be scrutinized thoroughly (see, e.g., Ref.~\cite{Carney:2022gse} for a bird’s-eye view of existing exclusions on such \dm model). Here, we demonstrate that next-generation space based GW detectors, such as Laser Interferometer Space Antenna (LISA) and Big Bang Observer (BBO)~\cite{2017arXiv170200786A,Crowder:2005nr},  are excellent testing grounds for strongly-interacting heavy non-annihilating DM.  Our proposal can be summarized as follows: Accumulation of strongly-interacting heavy non-annihilating DM inside binary stellar objects (symmetric Sun-like binaries, to be more specific) can transmute them into low mass (comparable with the progenitor masses) black hole binaries. These low mass BH binaries, commonly known as transmuted black hole (TBH) binaries, if sufficiently close, can emit quasi-monochromatic continuous GWs in their inspiral phase, and can be observed in the next-generation GW detectors, such as LISA and BBO. We theoretically estimate the occurrence rate-density of such TBH binaries whose progenitors are closely-spaced symmetric Sun-like binaries, and search these binaries in the future space-based GW detectors. Assuming a null detection of these TBH binaries with a year of observation time, we first calculate the upper limits on the occurrence rate-density, and by translating these upper limits on the occurrence rate-density, we provide novel (projected) constraints on heavy non-annihilating dark matter interactions.  The constraints derived in this work are subject to several astrophysical uncertainties but are complementary with the existing constraints from terrestrial direct searches and cosmological measurements.\\
	The rest of the work is organized as follows. In Sec.~\ref{2}, we briefly review the formation of low mass transmuted black holes. In Sec.~\ref{3}, we theoretically estimate the occurrence rate density of such transmuted black holes whose progenitors are closely-spaced symmetric Sun-like binaries. In Sec.~\ref{4}, we estimate the upper limit on the occurrence rate density by assuming a null detection of these low mass TBH binaries in future space-based GW detectors, such as LISA and BBO. In Sec.~\ref{5}, we use the upper limits on the occurrence rate density to derive the exclusion limits on the dark matter parameters (mass and its interaction cross-section with the nucleons). Finally, we culminate the paper with conclusions in Sec.~\ref{6}.
	\section{Formation of Low Mass Transmuted Black Holes}\label{2}
	In this section, we briefly review the transmutation process of binary stellar objects (for ease, we refer to binary stellar objects as stellar objects hereafter).\\
	Non-annihilating \dm particles from the Galactic halo that transit through a stellar object can be captured due to their collisions with stellar nuclei~\cite{1985ApJ...296..679P,Gould:1987ju,Gould:1987ir}. For sufficiently large DM-nucleon scattering cross-sections, the capture process becomes very effective. This is simply because, in this regime, \dm particles typically scatter many times while transiting through the stellar object, leading to a larger energy loss probability and almost all of the transiting \dm particles get captured. For heavy \dm, i.e., if the \dm mass $m_{\chi}$ is much heavier than the nuclei mass $m_A$ ($m_{\chi} \gg m_A$), which is of primary interest here, captured \dm particles sink toward the stellar core and settle into a small core-region. This leads to a huge number density of these captured \dm particles within the stellar core, which eventually collapses, followed by a small BH formation in the stellar core. This nascent BH, if not sufficiently light, can quickly devour the progenitor, transforming it into comparable mass BHs, which we refer as \textit{low-mass transmuted BHs}. In the following, we systematically go over the various stages of DM-induced transmutation of a stellar object (more specifically, Sun-like stars which are the progenitors in our analysis).\\
	\textbf{Capture:} To begin with, we first define the maximal capture rate as geometric capture rate $(C_{\rm{geo}})$, and it represents the total number of \dm particles that can pass through a stellar object. For a specific velocity distribution of the incoming \dm particles $f(u)$, the maximal capture rate is~\cite{Gould:1987ir}
	\begin{equation}
		C_{\rm{geo}} = \frac{\rho_{\chi}}{m_{\chi}} \pi R^2 \int \frac{f(u) du}{u} (u^2+v^2_{\rm{esc}})\,,
		\label{eqn:Cgeo}
	\end{equation}
	where $v_{\rm{esc}}$ is the escape velocity of the planetary systems, $R$ denotes the size of the stellar object, and $\rho_{\chi} = 0.4$\,GeV/cm$^3$ is the  Galactic \dm density in the solar neighborhood. Specifically, for a Maxwell-Boltzmann velocity distribution,  $C_{\rm{geo}}$ becomes 
	\begin{equation}
		C_{\rm{geo}} = \frac{\rho_{\chi}}{m_{\chi}} \pi R^2 \sqrt{\frac{8}{3 \pi}} \bar{v} \left(1+\frac{3 v^2_{\rm{esc}}}{2 \bar{v}^2}\right)\,,
	\end{equation}
	where $\bar{v} = 270 $\,km/s denotes the average velocity of the \dm particles in the Galactic halo. For Sun-like stars, the maximal capture rate is
	\begin{equation}
		C_{\rm geo}|_{\rm Sun-like}=1.3\times 10^{24}\, \textrm s^{-1}\left(\frac{10^6\, \rm{GeV}}{m_{\chi}}\right).
		\label{eq:Cgeosun}
	\end{equation}
	Depending on the DM-nucleon scattering cross-section and \dm mass, a certain fraction $(f_{\rm cap})$ of the \dm particles that transit get trapped, implying a capture rate $C=f_{\rm cap} C_{\rm{geo}}$~\cite{Bramante:2022pmn,Neufeld:2018slx,McKeen:2023ztq,Pospelov:2023mlz,Ema:2024oce}. For heavy \dm $(m_{\chi} \gg m_A$) and large DM-nucleon scattering cross-sections, which is of primary interest here, $f_{\rm cap}$  can even reach unity~\cite{Bramante:2022pmn}. Of course, for low DM-nucleon scattering cross-sections, capture becomes inefficient, and $f_{\rm cap}$ becomes extremely small (as shown in Fig.~\ref{fig:fcap}). We closely follow Ref.~\cite{Garani:2017jcj} to compute the capture fraction for our analysis, however, we have verified that using  $f_{\rm cap}$ from Ref.~\cite{Leane:2023woh} yields a slightly weaker result. We also note that enhancement of \dm capture rate due to close binary is negligible in this analysis~\cite{PhysRevLett.109.061301}.
	\begin{figure}
		
		\includegraphics[width=0.47\textwidth]{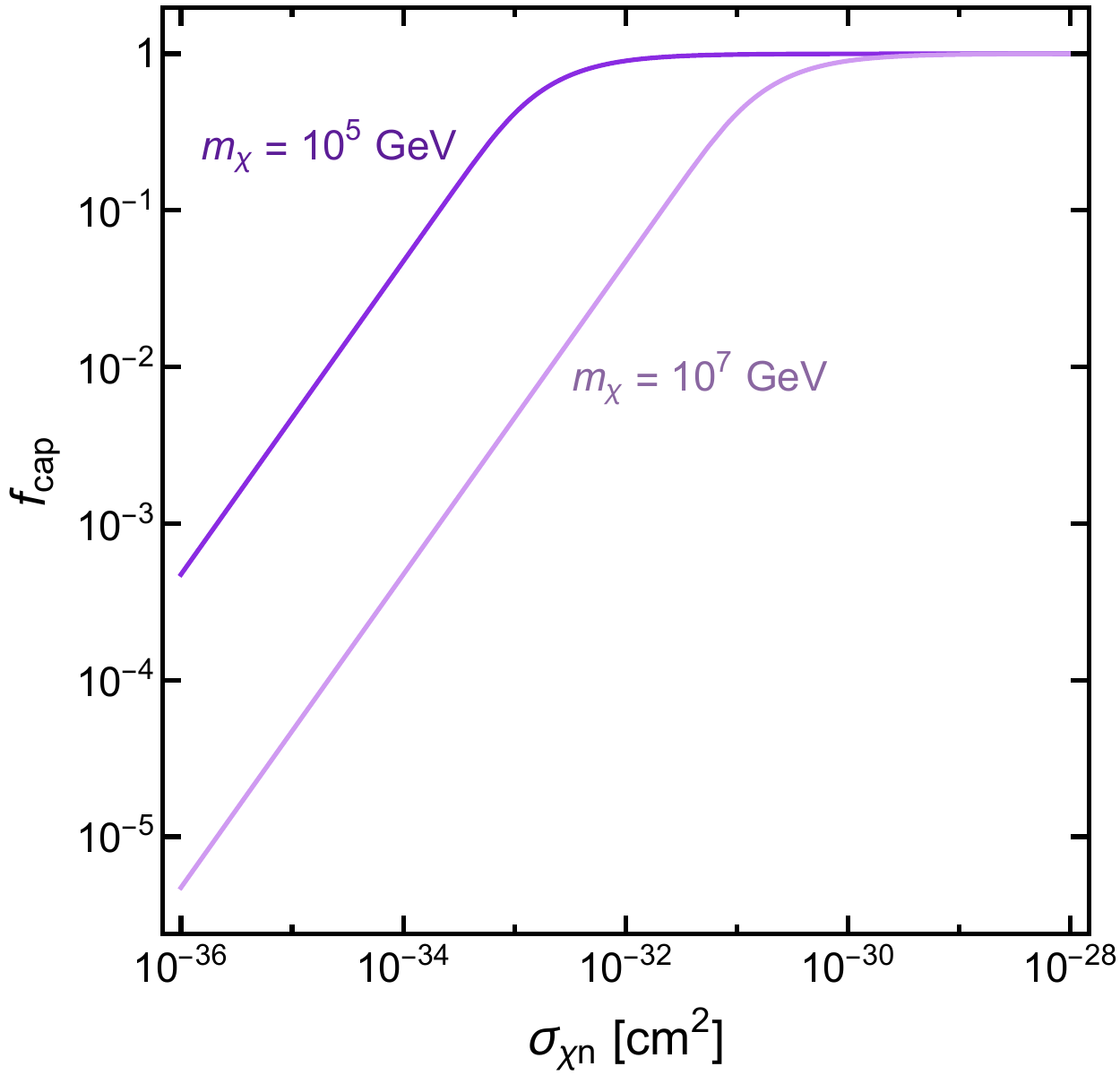}
		\caption{The capture fraction ($f_{\rm cap}=C/C_{\rm geo}$) is shown with the DM-nucleon scattering cross-section $(\sigma_{\chi n})$ for \dm masses of $10^5$ GeV and $10^7$ GeV. $f_{\rm cap}$ becomes very small for low $\sigma_{\chi n}$, and approaches unity for sufficiently large $\sigma_{\chi n}$. For heavier \dm, relatively large $\sigma_{\chi n}$ is needed in order to reach $f_{\rm cap} = 1$ as heavier \dm are harder to stop due to their larger kinetic energy.}
		\label{fig:fcap}
	\end{figure}
	\\\textbf{Spatial distribution inside the stellar volume:} After accumulation, \dm particles thermalize with the stellar nuclei via successive collisions. The timescale for thermalization depends strongly on the \dm-nucleon scattering cross-section: For large $\sigma_{\chi n}$, it occurs quickly, i.e., over a time much shorter than the stellar object lifetime~\cite{McDermott:2011jp,Kouvaris:2010jy,Kouvaris:2011fi,Bertoni:2013bsa,Garani:2018kkd,Acevedo:2020gro,Garani:2020wge}. The thermalized \dm particles then become spatially distributed in a way that depends crucially on their mass~\cite{Gould:1989hm,Leane:2022hkk}. For example, heavy \dm tends to shrink toward the stellar-core and stabilize into a tiny radius around the core, commonly known as the thermalization radius.  Quantitatively, for Sun-like stars, \dm particles of mass 10$^6$ GeV thermalize within a radius of  $r_{\rm th} \sim$ 91 km ($1.3\times 10^{-4}R_{\odot}$), which reduces further as $m^{-1/2}_{\chi}$ for heavier \dm.\\
	\textbf{Dark collapse and Black Hole formation:} The number of captured \dm particles grows linearly with time. As a consequence, stellar objects with cosmologically long lifetimes ($\sim$ Gyr) can accumulate an enormous number of \dm particles.  Quantitatively, for a \dm mass of $10^6$ GeV, and sufficiently high DM-nuclei scattering cross-section (say $10^{-28}$\,cm$^2$), $\mathcal{O}(10^{41})$ \dm particles accumulate inside a Sun-like star. When they concentrate within a $\sim$ 91 km radius, they give rise to a core-density of $\sim 6 \times 10^{25}$\,GeV cm$^{-3}$, around 26 orders of magnitude higher than the local Galactic \dm density. This huge core-density undergoes gravitational collapse once it exceeds the critical threshold value, and eventually leads to a micro-BH formation within the stellar core. The necessary and sufficient criterion for such a \bh formation has been extensively studied in the literature~\cite{Goldman:1989nd,Gould:1989gw,Bertone:2007ae,deLavallaz:2010wp,McDermott:2011jp,Kouvaris:2010jy,Kouvaris:2011fi,Bell:2013xk,Guver:2012ba,Bramante:2013hn,Bramante:2013nma,Kouvaris:2013kra,Bramante:2014zca,Garani:2018kkd,Kouvaris:2018wnh,Dasgupta:2020dik,Lin:2020zmm,Dasgupta:2020mqg,Acevedo:2020gro,Garani:2021gvc,Steigerwald:2022pjo,Singh:2022wvw,Ray:2023auh,Bhattacharya:2023stq,Bramante:2023djs}, and is essentially determined by the self-gravitating criterion and quantum degeneracy pressure. More specifically, BH formation occurs if 
	\begin{equation}
		N_{\chi}^{\rm BH}\rvert_{ t_{\rm{age}}} = C \times t_{\rm{age}} \geq \max \left[N^{\rm{self}}_{\chi}, N^{\rm{Cha}}_{\chi}\right]\,,
		\label{eq:Nchi}
	\end{equation}
	where $N_{\chi}^{\rm BH} \rvert_{ t_{\rm{age}}}$ denotes the total number of accumulated \dm particles in the stellar object throughout its lifetime ($t_{\rm age}$), $N^{\rm{self}}_{\chi}$ represents the self-gravitating criterion, and $N^{\rm{cha}}_{\chi}$ represents the quantum degeneracy pressure criterion. Below, we briefly describe these two criteria.\\
	Inside the thermalization volume, if the \dm density overcomes the baryonic density, then the \dm particles self-gravitate inside the core. This self-gravitating criterion ($N^{\rm{self}}_{\chi}$) is same for bosonic/fermionic \dm, and it only depends on the core-density $\rho_{\rm{core}}$ as well as core-temperature  $T_{\rm{core}}$ of the stellar objects~\cite{McDermott:2011jp} and the dependencies can be found as,
	\begin{eqnarray}
		N^{\rm{self}}_{\chi} &\sim& 2.8\times 10^{41} \left(\frac{\rho_{\rm{core}}}{154\, \rm{g/cm^3}}\right) \left(\frac{T_{\rm{core}}}{1.54 \times 10^7\,\rm{K}}\right)^{3/2} \nonumber \\ &\times& \left(\frac{10^6\,  \rm{GeV}}{m_{\chi}}\right)^{5/2}\,. 
		\label{eq:nchiself}
	\end{eqnarray}
	The quantum degeneracy pressure criterion ($N^{\rm{cha}}_{\chi}$) represents the maximum number of \dm particles beyond which \bh formation occurs. It solely depends on the spin of the \dm particles, and for bosonic (fermionic) \dm, it stems from the Heisenberg Uncertainty principle (Pauli exclusion principle)~\cite{McDermott:2011jp}. Quantitatively, 
	\begin{equation}
		N_{\chi}^{\rm Cha}|_{\rm boson}=9.5\times 10^{25}\left(\frac{10^6\,\rm GeV}{m_{\chi}}\right)^2\,,
	\end{equation}
	\begin{equation}
		N_{\chi}^{\rm Cha}|_{\rm fermion}=1.8\times 10^{39}\left(\frac{10^6\,\rm GeV}{m_{\chi}}\right)^3\,.
	\end{equation}
	From these numerical estimates, it is evident that, in our parameter range of interest, the transmutation criterion for Sun-like stars (which are the progenitors in our analysis) is determined by the self-gravitating criterion. This simply leads to the fact that the results derived from our analysis are applicable for both bosonic as well as fermionic \dm. The initial mass of the micro-BH formed at the stellar core $M_{\rm BH}$ can also be estimated as
	\begin{eqnarray}\label{bhmass}
		M_{\rm BH} &= m_{\chi} \times \max \left[N^{\rm{self}}_{\chi}, N^{\rm{Cha}}_{\chi}\right] = m_{\chi} \times N^{\rm{self}}_{\chi}&\label{eq:mbh}\\ & \sim 2.5 
		\times 10^{-10} M_{\odot} \left(\frac{10^6\,\rm{GeV}}{m_{\chi}}\right)^{3/2}\,\nonumber ,
	\end{eqnarray}
	where we take the solar core-density as 154 g/cm$^3$ and solar core-temperature as $1.54 \times 10^7$\,K~\cite{Acevedo:2020gro}.\\
	\textbf{Growth and Evaporation of newly-formed Black Holes:}
	The newly produced micro-BH at the center of the stellar core can accrete from the surrounding material and swallow the progenitor on a very short timescale (as compared to the stellar object's lifetime). It will also evaporate by emitting particles via Hawking radiation. For the time-evolution of the nascent BH, we can conservatively\footnote{This is conservative because we do not account for the accretion of newly incoming \dm particles by the micro-BH. } consider the baryonic matter accretion by the micro-BH~\cite{McDermott:2011jp,Kouvaris:2010jy}(first term) and the BH evaporation (second term)
	\begin{equation}
		\frac{dM_{\rm{BH}}}{dt} = \frac{4 \pi \rho_{\rm{core}} G^2M^2_{\rm{BH}}}{c^3_{s}} - \frac{P\,(M_{\rm{BH}})}{G^2M^2_{\rm{BH}}}\,,
	\end{equation}
	where $M_{\rm{BH}}$ denotes the mass of the newly formed BH and $c_s=\sqrt{T_{\rm{core}}/m_n}$ denotes the sound speed in the core of the stellar object, $T_{\rm{core}}$ is the core-temperature and $m_n$ is the nucleon mass. $\rho_{\rm{core}}$ denotes the core-density of the stellar object, $G$ is the gravitational constant, and $P\,(M_{\rm{BH}})$ denotes the Page factor~\cite{Page:1976df,MacGibbon:1991tj}. It is important to note that the Page factor properly accounts for gray-body corrections of the Hawking evaporation spectrum, as well as the number of Standard Model (SM) species emissions from an evaporating BH (BHs heavier than 10$^{17}$ g only emit massless particles, such as photons and neutrinos, whereas, lighter BHs emit massive SM particles too). In the classical black-body radiation limit, the Page factor evaluates
	to $1/\left(15360 \pi\right)$ and is commonly used in the literature. Considering the gray-body corrections and by accounting for the number of SM species emitted, the Page factor ranges from $1/(1135\pi)$ to $1/(74\pi)$~\cite{MacGibbon:1991tj,Arbey:2021mbl}.\\
	Since the accretion term scales as $M^2_{\rm{BH}}$, and the evaporation term scales as $1/M^2_{\rm{BH}}$, for low BH masses, evaporation dominates over the accretion process. Therefore, for sufficiently light micro-BHs, successful transmutation of the hosts do not occur. This sets a cutoff on the \dm mass that can be probed via transmutation as the mass of the micro-BH decreases with heavier \dm (Eq. \ref{eq:mbh}). For Sun-like systems, this cutoff mass is around $m_{\chi} \sim 10^{10}$ GeV~\cite{Ray:2023auh}.\\
	\textbf{Drift time and maximal possible scattering cross-section:}
	Transmutation of stellar objects does not occur at very large DM-nucleon scattering cross-sections. This is simply because, at very large DM-nucleon cross-sections, \dm particles lose a significant amount of energy in the outer shells of the stellar object, and therefore, may not reach the stellar core (or take a significantly longer time to reach the core). We estimate the drift time, i.e., the time required by the \dm particles to reach the stellar core by using the stellar density, temperature, and compositional profiles~\cite{Gould:1989gw,Bramante:2019fhi}
	\begin{equation}
		t_{\rm{drift}} = \frac{1}{G m_{\chi}} \sum_j \sigma_{\chi j} \int_{0}^{R} \frac{n_j(r) \sqrt{3A_jT(r)}}{\int_{0}^{r} d^3r^{\prime} \rho_j(r^{\prime})} dr\,,
	\end{equation}
	where $\sigma_{\chi j}$ denotes the DM-nuclei scattering cross-section, and is related to the DM-nucleon scattering cross-section via $\sigma_{\chi j} = \sigma_{\chi n}\,A^2_j \left(\mu_{\chi A_j}/\mu_{\chi n}\right)^2$. $A_j$ denotes the mass number of the $j$th nuclei, and $\mu_{\chi n}$ is the reduced mass of the DM-nucleon system. We set the ceilings of our results by demanding that $t_{\rm{drift}} \leq 1$\,Gyr. Quantitatively, for Sun-like systems (for both bosonic and fermionic \dm), it corresponds to~\cite{Acevedo:2020gro,Ray:2023auh}
	\begin{equation}
		\sigma_{\chi n} \leq 10^{-18}\, \textrm{cm}^2 \,\left(\frac{m_{\chi}}{10^{6}\,\rm{GeV}}\right)\,.
	\end{equation}
	\section{Occurrence Rate-density of low mass transmuted black holes}\label{3}
	Accumulation of strongly-interacting heavy non-annihilating \dm particles inside stellar objects can lead to the formation of comparable mass TBHs. These low-mass TBHs, while in binaries, can emit GWs as they inspiral toward each other, and could be detected in the next-generation space-based GW detectors, such as LISA. As a concrete example, binaries with component masses of 1$M_{\odot}$ and separation of 4$R_{\odot}$ emit GWs at a frequency of $3.5 \times 10^{-5}$ Hz, potentially detectable by LISA (as shown in Fig.~\ref{fig:rorb-vs-fgw}). Of course, the initial separation of these progenitors binaries can vary, which would then lead to GW emission at a different frequency. For wide binaries (binaries with larger separation), the GW frequency decreases and eventually falls outside of the LISA sensitivity band. Quantitatively, for symmetric Sun-like binaries (which are the progenitors in our analysis), if the orbital separation exceeds 9.5$R_{\odot}$, the GW frequency becomes $< 10^{-5}$ Hz and falls outside the LISA sensitivity (as shown in Fig.~\ref{fig:rorb-vs-fgw}). This implies that in our analysis we only consider GW emission from Sun-like binaries with orbital separations between $4R_{\odot} - 9.5R_{\odot}$. Note that we choose Sun-like systems as our progenitors because (1) they have much larger sizes as compared to the planets ensuring maximum \dm accumulation (Eq.~\ref{eqn:Cgeo}), and (2) stars form binaries much more easily than planets do.\\
	In the following, we theoretically estimate the occurrence rate-density of such TBH binaries whose progenitors are closely-spaced symmetric Sun-like binaries. First, we define the key quantity for this estimation: transmutation time $(\tau_{\rm trans})$ which dictates the total time required in order to have a successful transmutation of the progenitors. $\tau_{\rm trans}$ is essentially a sum of two timescales, where the first timescale represents the time required to form a micro-BH inside the stellar core, and the second timescale represents the time required to swallow the progenitor by the newly formed micro-BH. In our parameter range of interest, the first timescale, which depends on the \dm mass as well as the DM-nucleon scattering cross-section, always exceeds the second timescale, which only depends on the \dm mass. Of course, transmutation time $(\tau_{\rm trans})$ also depends on the ambient \dm density which we take as 0.4 GeV/cm$^3$ as our region of interest resides in the solar neighborhood.\\
	We consider a closely-spaced, sun-like binary system, which is formed at time $t_f$. For successful transmutation, the time required for transmutation has to be shorter than the available time, implying,
	\begin{equation}
		\tau_{\rm trans} \leq (t_0 - t_f)\,,
	\end{equation}
	where $t_0 = 13.79$ Gyr denotes the current age of the Universe. Clearly, only a fraction of these binaries that satisfy the above criterion will undergo a transmutation.  Therefore, the occurrence rate density of such \tbh binaries (whose progenitors are closely-spaced symmetric Sun-like binaries) can be written as
	\begin{align}
		R_{\rm TBH} & \propto \frac{1}{t_{\rm obs}}\int_{t_*}^{t_0} dt_f\,\lambda\,\frac{d\rho^*}{dt}[t_f]  \times
		\Theta\left[t_0-t_f-\tau_{\rm trans}(m_{\chi}, \sigma_{\chi n})\right]\,,
		\label{eq:Rtbh}  
	\end{align}
	where $t_{\rm obs} = 1$ year denotes the observation time and $t_* = 4.9 \times 10^8$\,year $(z_*=10)$~\cite{Taylor:2012db} denotes the earliest binary formation time. $\lambda$ denotes the fraction of stellar mass in binaries and $\frac{d\rho^*}{dt}[t_f]$ denotes the cosmic star formation rate density which we take as Madau and Dickinson star formation rate density. For the normalization of Eq.(\ref{eq:Rtbh}), we use the total number of progenitors (closely-spaced symmetric Sun-like binaries) within our volume of interest and assume a uniform spatial distribution of the progenitors in our Galaxy.\\ 
	In the following, we provide a brief estimate of the normalization criterion that has been used in our analysis. Our Galaxy contains almost $200\times 10^9$ stars, and of which about 20\% are Sun-like stars, implying $40\times 10^9$ Sun-like stars in our Galaxy~\cite{Popov:2007bhy}. Among them, a certain fraction can form a closely-spaced symmetric binary, which we denote as $\alpha$. This implies the total number of progenitors (closely-spaced symmetric Sun-like binaries) in our Galaxy is $40\times 10^9 \alpha$, and by assuming a uniform spatial distribution of the progenitors in our Galaxy, this leads to the normalization condition for $R_{\rm TBH}$ being $\frac{40\times 10^9\alpha}{\frac{4}{3} \pi (15 \rm{kpc})^3}$ for an observation time of one year. Here, we note that the closely-spaced symmetric binary fraction $\alpha$ is rare and it has large astrophysical uncertainties (see e.g., Ref.~\cite{2019ApJ...875...61M} and references therein). Quantitatively, for Sun-like ($M\approx0.5-1.6 M_{\odot}$) binaries with solar metallicity, the close binary fraction (orbital separation $\lesssim$ 100$R_{\odot}$) is reported to be smaller than 0.05~\cite{2019ApJ...875...61M}. Therefore, our choices of $\alpha$, which are used in this analysis, are well justified. As these estimates are subject to astrophysical uncertainties, we take $\alpha$ as a free parameter and vary it over a wide range of values ($\alpha\sim 10^{-3}-10^{-11}$), which are consistent with the literature.
	\begin{figure}
		\includegraphics[width=0.47\textwidth]{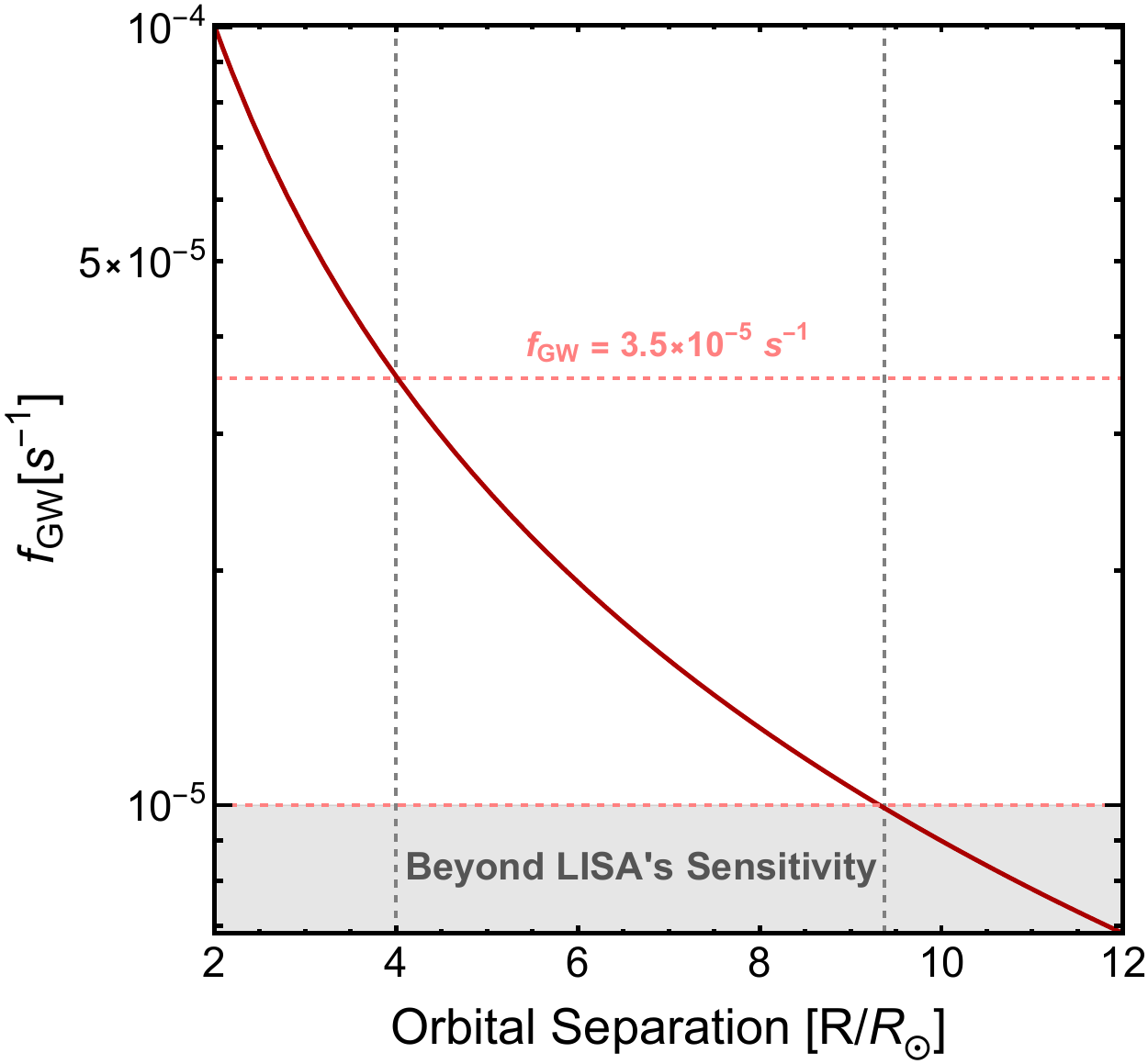}
		\caption{GW frequency $(f_{\rm GW})$ is shown (solid red line) as a function of the orbital separation for a symmetric Sun-like binary. We probe binaries with orbital separations ranging from 4$R_{\odot}$ to 9.5$R_{\odot}$, which corresponds to the frequency range of $(10^{-5}$ to $3.5 \times 10^{-5})$ Hz. Above an orbital separation of 9.5$R_{\odot}$, the frequency falls outside the LISA band.}
		\label{fig:rorb-vs-fgw}
	\end{figure}
	\section{Continuous-wave probes of low mass transmuted black holes}\label{4}
	In the previous section, we estimated the occurrence rate density of TBH binary systems that originated from closely-spaced sun-like stars. We now must calculate the expected sensitivity of space-based \gwh detectors toward these TBH systems that slowly inspiral toward one another, specifically the expected rate density constraints on such systems, and compare that to \cref{eq:Rtbh} for different choices of the fraction of stars that form in binaries $\alpha$.

	The slow inspiral of two objects orbiting around each other can be described as a continuous wave if they are either far enough away from each other or at low enough frequencies. When these conditions are met, the time/frequency ``chirp'' characteristic of detected binary inspirals can be approximated as a slow linear increase of the \gwh frequency over time.\\
	Quantitatively, consider the rate of change of \gwh frequency over time, $\dot{f}_{\rm GW}$, of an inspiraling system sufficiently far from merger \cite{maggiore2008gravitational}
	\begin{eqnarray}
		\dot{f}_{\rm GW}&=&\frac{96}{5}\pi^{8/3}\left(\frac{G\mathcal{M}}{c^3}\right)^{5/3} f_{\rm GW}^{11/3} \nonumber \\
		&\simeq& 10^{-23}\text{ Hz/s} \left(\frac{\mathcal{M}}{0.87M_\odot}\right)^{5/3}\left(\frac{f_{\rm GW}}{3.5\times 10^{-5}\text{ Hz}}\right)^{11/3} \
		\label{eqn:fdot_chirp}
	\end{eqnarray}
	where $\mathcal{M}\equiv\frac{(m_1m_2)^{3/5}}{(m_1+m_2)^{1/5}}$ is the chirp mass, $m_1$ and $m_2$ are the component masses, and $f_{\rm GW}$ is the \gwh frequency. Its integral is
	\begin{equation}
		f_{\rm GW}(t)=f_0\left[1-\frac{8}{3}\frac{\dot{f}_{\rm GW}}{f_0}(t-t_{\rm ref})\right]^{-\frac{3}{8}}~,
		\label{eqn:powlaws}
	\end{equation}
	where $f_0$ is the \gwh frequency at a reference time $t_{\rm ref}$.
	From Eq.~(\ref{eqn:fdot_chirp}), we can see that, for a system with $m_1=m_2=1M_\odot$, and at low-enough frequencies $(f_0 \sim 10^{-5}$\,Hz), the second term in Eq.~(\ref{eqn:powlaws}) will be much less than 1, and can be binomially expanded to
	\begin{equation}
		f_{\rm GW} = f_0+\dot{f}_{\rm GW}(t-t_{\rm ref})\,,
		\label{eqn:fcw}
	\end{equation}
	which describes quasi-monochromatic \cws.\\
	For a particular detector, we can compute the minimum detectable signal amplitude $h_0$ in a matched-filtering search for a quasi-monochromatic \gw with a \snr $\rho=1$ (corresponding to the best possible sensitivity that we could have) \cite{maggiore2008gravitational,Astone:2014esa}. Matched filtering correlates a model waveform with the data, and in the case of a purely sinusoidal signal,  it is simply a fast Fourier transform of the data neglecting the antenna patterns of the detectors, which induce $\mathcal{O}(1)$ changes
	\begin{equation}
		h_{0,\rm min} = 2\sqrt{\frac{S_n(f_{\rm GW})}{\Tobs}},
		\label{eqn:h0mf}
	\end{equation}
	where $S_n(f)$ is the noise power spectral density as a function of frequency $f$, and $\Tobs$ is the observation time. Here, we use the sensitivity curves for LISA and BBO given in \cite{Moore:2014lga}.\\ 
	The choice of which signal-to-noise ratio to use in this work is somewhat arbitrary. In a real search, a threshold is set on $\rho$ to ensure a certain false alarm rate, leading to values of $\sim 8-10$ \cite{maggiore2008gravitational} depending on (1) how much computational power is available, (2) how many outliers one would like to follow up, and (3) how large of a parameter space one is exploring. We do not know, in practice, how a search in LISA data will be performed, and thus quote the best, i.e. ``nominal'', sensitivity, defined as a signal with $\rho=1$, meaning a signal that is detectable at the level of the noise. We note, however, that the actual sensitivity will certainly be worse than what we quote; however, the reduction in sensitivity in practice is much smaller than, say, the uncertainty on the parameter $\alpha$.\\
	We can also write an equation for the distance reach to binaries at different frequencies \cite{maggiore2008gravitational}
	\begin{align}
		d(\fgw)&=\frac{4}{h_0}\left(\frac{G \mathcal{M}}{c^2}\right)^{5/3}\left(\frac{\pi \fgw}{c}\right)^{2/3} \label{eqn:d} \\
		&= 0.27\text{ pc} \left(\frac{3.66\times 10^{-19}} {h_0}\right)\left(\frac{\mathcal{M}}{0.87M_\odot}\right)^{5/3}\nonumber \\ &\times \left(\frac{f_{\rm GW}}{3.5\times 10^{-5}\text{ Hz}}\right)^{2/3} \nonumber
	\end{align}
	From the distance reach, and following the procedure in \cite{Miller:2021knj}, we can calculate the number of binaries that we would detect in a given observing run by multiplying the space-time volume $\avgVT$ (which, for nearby binaries, is the volume of a sphere times the duration of the sources) by the expected formation rate density $\mathcal{R}$, assuming that the binaries are uniformly distributed over our Galaxy \cite{Miller:2021knj}
	\be
	N_{\rm bin}(\fgw) \simeq \avgVT\mathcal{R} = \frac{4}{3}\pi [d(\fgw)]^3\mathcal{R}T(\fgw),\label{eqn:Nbin}
	\ee
	{\( T\) is the time over which we integrate the binary systems' frequency evolutions: \(T = \mathrm{max}(t_\mathrm{obs},\Delta T)\), and \(\Delta T\) is how long the binary system spends in the frequency range $[f,f+\delta f]$, which can be obtained by inverting Eq.~(\ref{eqn:powlaws})} 
	\begin{align}
		\Delta T = \frac{5}{256}\pi^{-8/3}\left(\frac{c^3}{G\mathcal{M}}\right)^{5/3} \left[\fgw^{-8/3}-(\fgw+\delta f)^{-8/3 }\right].
		\label{eqn:deltaT}
	\end{align}
	When $\Delta T$ exceeds $\Tobs$, the number of detectable sources is dominated by those between the frequencies $(\fgw,\fgw+\delta f)$, including those that began emitting \gws well before the observation run. Here, $\delta f=1/\Tobs$ is the resolution in frequency and indicates that the frequency of the \gwh signal does not vary by more than one frequency bin during $\Tobs$.
	Essentially, when $\Delta T>\Tobs$, we allow for the enhancement of the signal during the observation time due to the fact that many possible systems inspiraling forever would be emitting \gws at the same frequency. In other words, sources coming from anywhere in the sky, with $\dot{f}_{\rm GW}\sim \mathcal{O}(10^{-23})$ Hz/s, are \emph{indistinguishable} from each other at a given frequency, and thus, their powers will add. These sources are indistinguishable because the frequency shift induced by the relative motion of the earth with respect to the source, $\Delta f_{\rm Doppler}=10^{-4}\fgw\sim \mathcal{O}(10^{-9})$ Hz, and the shift caused by the spin-up of the source in $\Tobs$, $\Delta f_{\dot{f}}=\dot{f}_{\rm GW}\Tobs \sim \mathcal{O}(10^{-16})$ Hz, are both well within the frequency resolution $\delta f$ of a matched-filtering search over $\Tobs$, i.e. $\delta f\gg\Delta f_{\rm Doppler}$ and  $\delta f\gg\Delta f_{\dot{f}}$.\\
	Here, we also impose that $\Delta T$ cannot exceed the time at which the first stars begin to form binaries, i.e., $\Delta T \lesssim t_0/10$.\\
	After calculating $N_{\rm bin} (\fgw)$, we sum the number of binaries emitting \gws at each frequency
	\be
	N_{\rm bin}^{\rm tot} = \sum_{ i} N_{\rm bin} (f_{\text{GW},i})~ < 1
	\label{eqn:ntot}
	\ee
	where we have required $N_{\rm bin}^{\rm tot}<1$ to be in the case in which we do not observe these systems in the future, in order to get the strongest constraint on $\mathcal{R}$.
	Then, we solve for $\mathcal{R}$, which provides an upper limit on the formation rate densities of such binary planet systems as a function of their chirp masses
	\be
	\mathcal{R}=\frac{3}{4\pi}\left(\sum_i T(f_{\text{GW},i})d(f_{\text{GW},i})^3\right)^{-1}.
	\label{eqn:ratedenssolved}
	\ee
	In order to perform the above computation, we must impose a minimum separation between the two objects, and this condition leads to the maximum \gwh frequency at which we could detect these systems. On the other hand, the minimum frequency arises from the detector's sensitivity band. Thus, we need to restrict the orbital radius to be
	\begin{equation}
		\rorb>2(R_1+R_2),\label{eqn:rorb-cond}
	\end{equation}  
	where $R_1$ and $R_2$ are the radii of the two planets, $\rorb=\left(\frac{G m}{\pi^2 \fgw^2}\right)^{1/3}$ and $m=m_1+m_2$.\\
	While we can in theory calculate the rate densities for any given $m_1$ and $m_2$, here we restrict ourselves to solar-mass binaries, i.e. $R_1=R_2=R_\odot$ and $m_1=m_2=\msun$. 
	To perform this computation, we sum over the luminosity distance reaches as a function of frequency only at frequencies for which Eq.~(\ref{eqn:rorb-cond}) is satisfied. In Fig. \ref{fig:rorb-vs-fgw}, we show the orbital separation of the binary as a function of the \gwh frequency, for an equal solar-mass binary. The maximum frequency for which the condition in Eq.~(\ref{eqn:rorb-cond}) is satisfied, is drawn as a horizontal line (dashed red).\\ 
	Applying the procedure described above for an equal solar-mass binary, and assuming that we do not find any such systems in future space-based detectors, such as LISA and BBO, we obtain a projection for the upper limit on the occurrence density of such systems 
	\begin{equation}\label{limit}
		\mathcal{R}\leq
		\begin{cases}
			19\text{ \invkpccubedyr}, & \text{LISA}, \\
			8.22\times 10^{-6} \text{ \invkpccubedyr}, & \text{BBO}.
		\end{cases}
	\end{equation}
	If we, instead, assume $\Tobs=5$ years, a possible lifetime for space-based detectors, our rate density upper limits would become $8.74\text{ \invkpccubedyr}$ and $3.65\times10^{-6}\text{ \invkpccubedyr}$ for LISA and BBO, respectively, which are stronger than those in \cref{limit}. Note that our estimates of the rate densities depend both on the distance reach and $\Delta T$, as $\Delta T \gg \Tobs$ at all frequencies considered here. From \cref{eqn:h0mf} and \cref{eqn:d}, we can see that the distance reach scales with $\sqrt{\Tobs}$, while  $\Delta T$ depends on $\Tobs$ via the frequency range through which we consider the source to sweep. The interplay between these two factors -- longer observation times improve $d$, but also shrink $\Delta T$, since the frequency spread is smaller -- results in a nontrivial change of the rate density upper limits when changing $\Tobs$.\\
	This is our main result in Sec.~\ref{4}, and with these rate density upper limits, we now compute the projected constraints on strongly-interacting heavy non-annihilating particle \dm parameters in Sec.~\ref{5}.
	\section{Projected constraints on dark matter parameters}\label{5}
	\begin{figure*}
		\centering
		\includegraphics[width=0.42\textwidth]{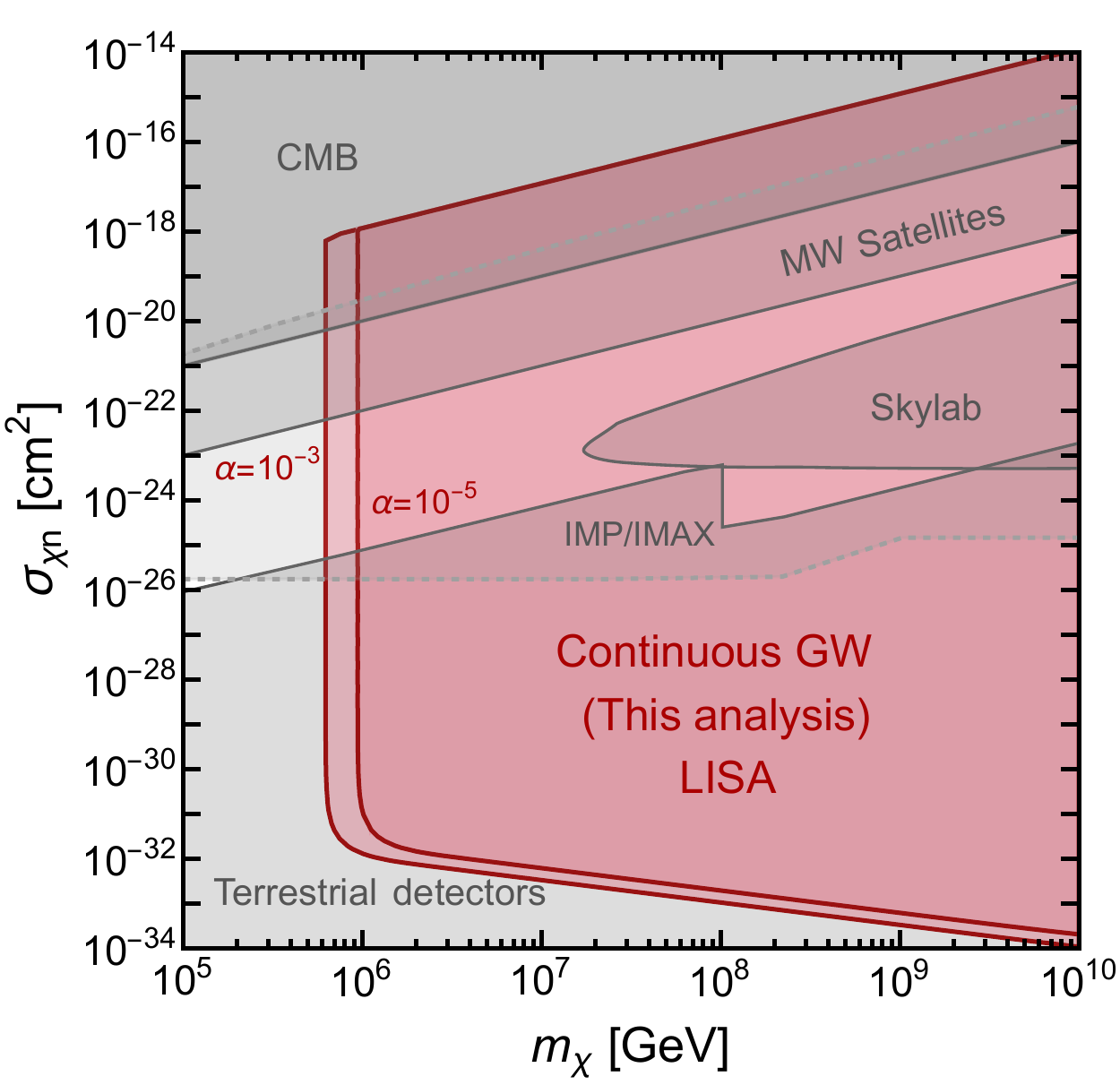}
		\hspace{1 cm}
		\includegraphics[width=0.42\textwidth]{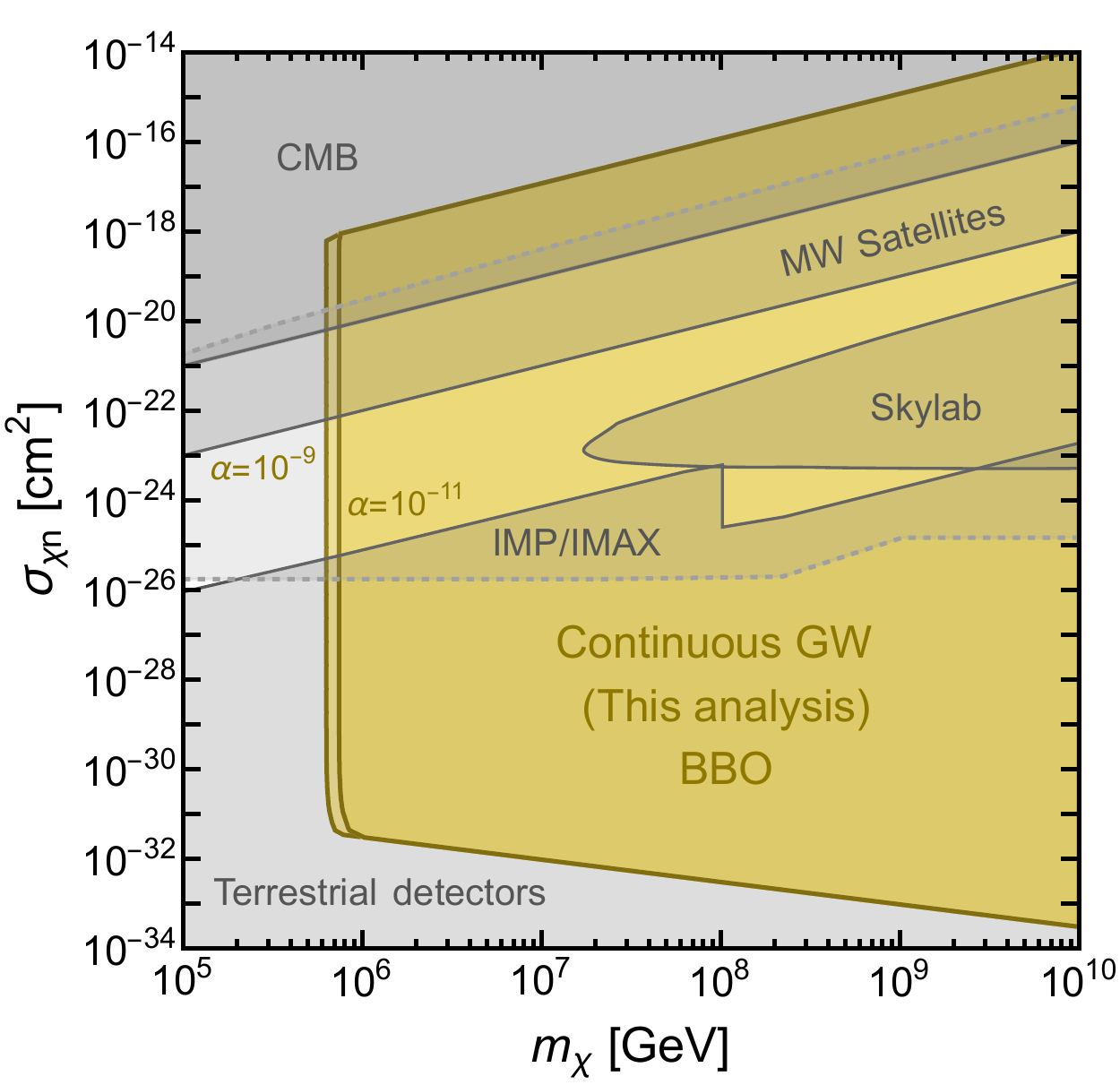}
		\caption{Projected constraints on \dm mass and its interaction cross-section with the nucleons from continuous GW searches with the future space-based detectors, such as LISA (left) and BBO (right). The constraints are derived by assuming a null detection of the transmuted black binaries (whose progenitors are symmetric Sun-like binaries) with a year of observation time and apply for both bosonic/fermionic \dm particles. In the (left) right panel, (red) yellow-shaded regions denote the constraints derived in this work, whereas, the gray-shaded regions denote the existing constraints (see text). Our constraints crucially depend on $\alpha$, closely-spaced symmetric binary fraction, and we show our constraints for $\alpha = 10^{-3}$ ($\alpha = 10^{-9}$) and $\alpha = 10^{-5}$ ($\alpha = 10^{-11}$) for LISA (BBO), consistent with the literature (see e.g., Ref.~\cite{2019ApJ...875...61M} and references therein). We note that for $\alpha \leq 10^{-6}$ ($\alpha \leq 10^{-12}$), we do not obtain any exclusion on the \dm parameters for LISA (BBO).}
		\label{fig:constraint}
	\end{figure*}
	For \textit{conservative} constraints on particle \dm parameters (\dm mass and its interaction strength with nucleons), we compare the theoretical occurrence rate density of \tbh binaries (whose progenitors are symmetric closely-spaced Sun-like binaries) (Eq.~\ref{eq:Rtbh}) to the corresponding upper limits obtained from continuous \gwh searches (Eq.~\ref{limit}). More specifically, our projected constraints are simply derived from $R_{\rm TBH} \leq 19\text{ \invkpccubedyr}$ (for LISA) and $R_{\rm TBH} \leq 8.22 \times 10^{-6}\text{ \invkpccubedyr}$ (for BBO).\\
	In Fig.~\ref{fig:constraint}, we show the projected exclusion limits, assuming non-observation of continuous gravitational waves from stellar binaries, on $\{m_{\chi}, \sigma_{\chi n}\}$ obtained from future space-based detectors like LISA (left) and BBO (right) for strongly-interacting heavy non-annihilating \dm particles. In the left (right) panel, red (yellow) shaded regions are the projected exclusions from our analysis, whereas, the gray shaded regions are the existing exclusions from a variety of searches.  For our analysis, we assume an observation time of one year, and with a larger observation time the constraints get even stronger. Note that, the constraints  in  Fig.~\ref{fig:constraint} applies for both bosonic as well as fermionic \dm. This is simply because for Sun-like stars, the transmutation criterion is essentially determined by $N^{\rm{self}}_{\chi}$, and this is independent of the spin of the \dm particles.  Our constraints also crucially depend on the closely-space symmetric binary fraction $\alpha$, which has large astrophysical uncertainties~\cite{2019ApJ...875...61M}. Here, we show our constraints for some reasonable choices of $\alpha$ ($\alpha = 10^{-3}$, $10^{-5}$ for LISA \& $\alpha = 10^{-9}$, $10^{-11}$ for BBO), which are quite consistent with the literature~\cite{2019ApJ...875...61M}. So, it is evident that even if the closely-spaced symmetric binary fraction ($\alpha$) is rare, the constraints obtained from continuous GW searches are already competitive with the existing exclusion limits, demonstrating that future space-based GW detectors are ideal laboratories to probe heavy non-annihilating dark matter.\\
	We also consider  the case in which we use \emph{semi-coherent} estimates of sensitivity for a future LISA search for transmuting Sun-like binaries. Semi-coherent methods break the data of duration $\Tobs$ into smaller chunks of length $\TFFT$ that are analyzed coherently (keeping the phase information) and combined incoherently (without the phase information). Such methods may be more realistic to use in LISA data analysis, since the generation of waveforms takes an immense amount of computational time. The tradeoff for computationally efficiency, however, is a reduction in sensitivity. More specifically, as shown in \cite{Astone:2014esa}, the sensitivity loss of using semi-coherent methods with respect to matched filtering is a factor of a few, $\sim 3$, depending on the choice of $\TFFT$. In this case, the minimum amplitude given in \cref{eqn:h0mf} is approximately a factor of a few ($\sim 3$) higher than that obtained from matched filtering, meaning that the rate densities estimated in \cref{limit} would increase by a factor of $\sim 27$. In Fig.~\ref{fig:matched_vs_semi}, we compare our results using matched filtering technique (solid red, same as left panel of Fig.~{\ref{fig:constraint}}) as well as semi-coherent analysis (dashed red) for a fixed close-binary fraction $(\alpha = 10^{-3})$. We show that both these techniques yield similar constraints on heavy non-annihilating dark matter interactions. \\
	In the following we briefly describe the existing exclusion limits which are shown in gray shaded regions. Cosmological constraints, labeled as, ``CMB" and ``MW Satellites" denote exclusions obtained from Planck measurements of temperature and polarization anisotropy of the cosmic microwave background~\cite{Gluscevic:2017ywp,Boddy:2018wzy} and  observations of Milky Way satellite galaxies~\cite{DES:2020fxi,Nadler:2019zrb}, respectively.  Large panels of etched plastic, placed aboard the Skylab Space Station, also provide significant exclusion limits on DM-nucleon interactions, labeled as ``Skylab"~\cite{Starkman:1990nj,Bhoonah:2020fys,Wandelt:2000ad}. Constraints labeled as ``Terrestrial detectors" represents a bird's-eye view of the existing constraints from underground, surface, and high altitude detectors, and is taken from~\cite{Kavanagh:2017cru, Digman:2019wdm,Carney:2022gse}. Other astrophysical constraints, such as, disk stability~\cite{Starkman:1990nj}, interstellar gas cooling~\cite{Chivukula:1989cc}, as well as Galactic Center gas-cloud heating~\cite{Bhoonah:2018wmw,Bhoonah:2020dzs}; terrestrial constraints, such as MAJORANA demonstrator at the Sanford underground research facility~\cite{Clark:2020mna}, DEAP-3600 detector at SNOLAB~\cite{DEAPCollaboration:2021raj}, a shallow-depth experiment performed at the University of Chicago~\cite{Cappiello:2020lbk}, and Rocket-based X-ray Quantum Calorimetry (XQC) experiment~\cite{Erickcek:2007jv} do not cover any additional parameter space, and hence are not shown for clarity. Exclusion limits obtained from the mere existence of the Sun, and other solar-system planets cover a similar parameter space~\cite{Ray:2023auh,Acevedo:2020gro}, and are also not shown for clarity. Finally, constraints obtained from cosmic ray silicon detector satellite (IMP7/8), and balloon-borne experiment (IMAX) are shown in gray (thin) dashed line as they are not based on detailed analyses in peer-reviewed papers~\cite{Wandelt:2000ad}.\\
	In this analysis, we consider Sun-like systems as our progenitors because in the strongly-interacting regime, the Sun captures a lot more \dm particles as compared to other planetary bodies, and more importantly, the binary formation for stars is much more favorable than planetary bodies (for planets, $\alpha$ is significantly small). The exclusion limits in Fig.~\ref{fig:constraint} can be understood qualitatively from the following. For lighter \dm, transmutation criterion is not attainable as the total number of captured \dm particles inside the stellar core $(\sim 1/m_{\chi})$ is not sufficient for transmutation $(\sim 1/m^{5/2}_{\chi})$. This sets the sharp vertical cutoff in lighter \dm mass (around $m_{\chi} \sim 10^6$\,GeV). For heavier \dm, transmutation is more favourable, but the mass of micro-BH decreases with an increase in \dm mass (Eq.~\ref{bhmass}). As a result,  the micro-BH takes a substantially
	longer time to consume the host (much longer than the age of the stellar object), and Hawking radiation becomes dominant over the accretion processes.  The combination of these two
	effects prevents successful transmutation and provides the vertical cutoffs at the heavier \dm masses (around $m_{\chi} \sim 10^{10}$\,GeV). For sufficiently low DM-nucleon scattering cross-sections, the capture fraction $(f_{\rm cap})$ decreases, and eventually the total number of captured \dm particles inside the stellar core is not sufficient for \bh formation. Because of this inefficient capture with low DM-nucleon scattering cross-sections, we cannot probe arbitrarily low DM-nucleon scattering cross-sections, leading to the lower boundary of our exclusion regions. Very large DM-nucleon scattering cross-sections are also not probed via transmutations as the drift time of the \dm particles becomes substantially longer. In Fig.~\ref{fig:constraint}, we show the ceilings of our results by demanding that $t_{\rm{drift}} \leq 1$\,Gyr, which corresponds to $\sigma_{\chi n} \leq 10^{-18}\, \textrm{cm}^2$ for $m_{\chi} = 10^{6}$\,GeV  and linearly increases with heavier \dm mass. We also note that for $\alpha \leq 10^{-6}$ ($\alpha \leq 10^{-12}$), we do not obtain any exclusion on the \dm parameters for LISA (BBO).
	\section{Summary \& Conclusions}\label{6}
	In this work, we propose a novel way to probe strongly interacting heavy non-annihilating \dm interactions with ordinary baryonic matter. Such \dm model is hard to probe in the conventional direct detection experiments because of their tiny fluxes, and demands new techniques.  We show that next-generation space-based GW detectors, such as LISA, are ideal testing room for such \dm model. We propose gradual accumulation of heavy, non-annihilating \dm in the ordinary stars (Sun-like) can lead to comparable mass black holes, and these low mass BHs, while in binaries, can emit quasi mono-chromatic GWs as they inspiral toward each other. For closely-spaced symmetric Sun-like binaries (which are the progenitors in our analysis) the frequency of these GWs is $\mathcal{O} (10^{-5})$ Hz, detectable by future space-based GW detectors, such as LISA. So, we search these closely-spaced  solar-mass transmuted binaries in LISA, and estimate model-independent upper limits on their rate densities by assuming a null detection of these binaries with a year of observation time. Finally,  we translate these limits to put significant (projected) exclusions on heavy non-annihilating \dm interactions, demonstrating the potential of space-based GW detectors as probes of heavy non-annihilating \dm.\\
	The novel exclusion limits obtained in the analysis apply to bosonic as well as fermion dark matter particles and cover a mass window from $10^6$ GeV to $10^{10}$ GeV. The lower mass cut-off arises from the fact that the transmutation criterion is harder to achieve for light \dm particles, whereas, the higher mass cutoff stems from the fact that the newly produced micro-BH is inefficient for causing a successful transmutation.  The exclusion limits also depend on the closely-spaced symmetric binary fraction of the progenitors $(\alpha)$, which is rare and currently uncertain. However, for reasonable choices of the closely-spaced symmetric binary fraction of the progenitors, we find that the exclusion limits derived in this analysis are competitive with the existing constraints obtained from cosmological as well as direct searches. Note that, while we have focused on symmetric Sun-like stars in this analysis, our results can easily be generalized for binaries with any component masses, by correctly accounting for the closely-spaced binary fraction, which is even rarer for planetary bodies. Therefore, our work 
	opens up a new window to probe heavy non-annihilating \dm interactions by using continuous GW techniques.\\ 
	We must also mention that our results are sensitive to whatever the true low-frequency cutoff of the future space-based detectors will be.  If future space-based detectors cannot reach such a low frequency, it would be difficult to probe solar-mass binaries in this way, and instead, we would have to consider much heavier systems, for which we need to estimate the close binary fraction correctly. Such a task depends heavily on uncertain astrophysics and can be looked at in future work. Additionally, other future space-based detectors, such as Taiji and TainQin \cite{Hu:2017mde,Luo:2015ght}, may also be useful in probing this kind of \dm.\\
		\begin{figure}
		\includegraphics[width=0.45\textwidth]{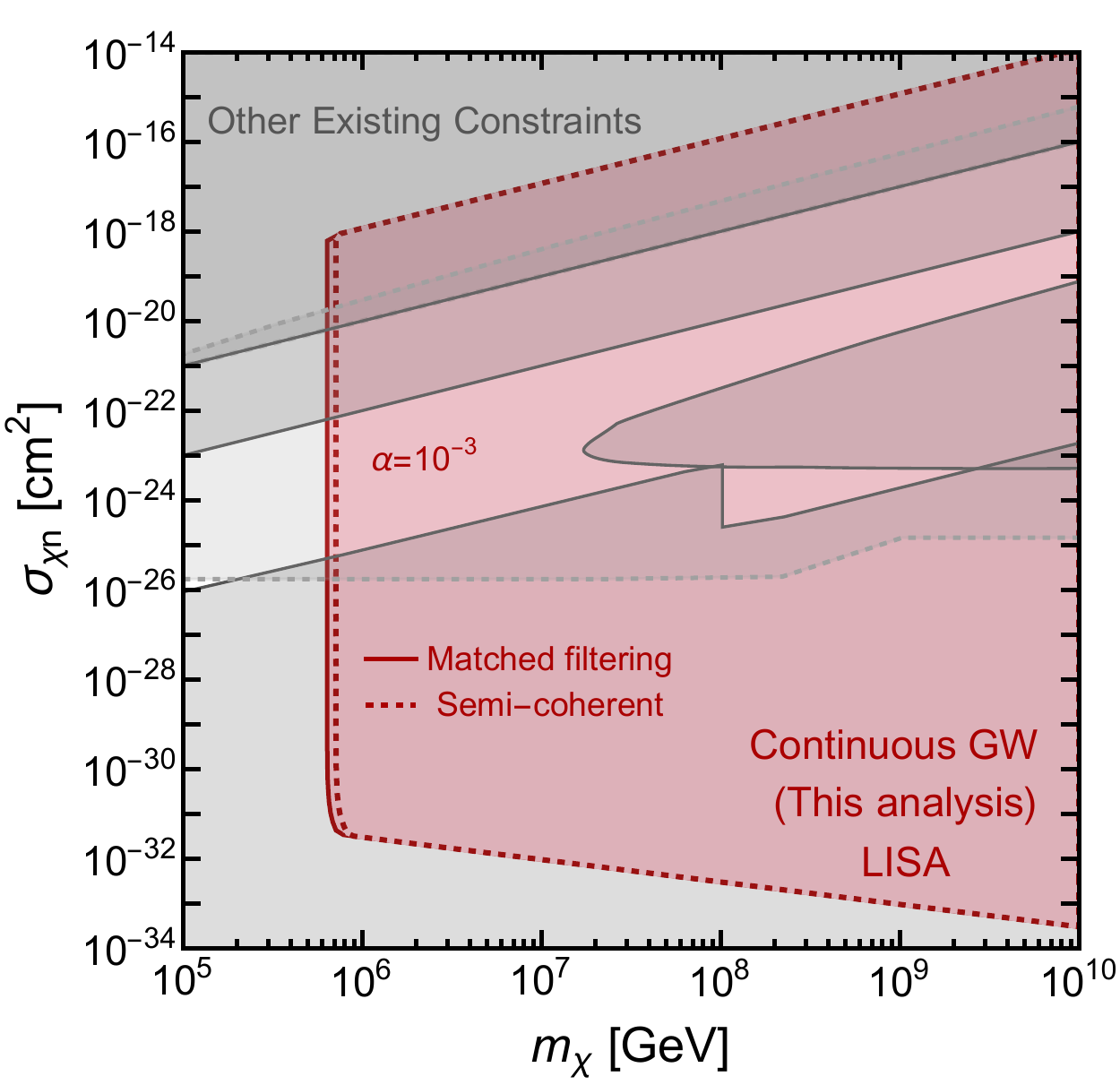}
		\caption{Projected constraints on \dm mass and its interaction cross-section with the nucleons from continuous GW searches with the future space-based detector, LISA. We take the close-binary fraction  $\alpha = 10^{-3}$, and compare the results for matched-filtering (solid red) and semi-coherent (dashed red) search techniques. The gray-shaded regions are the same as Fig~{\ref{fig:constraint}} and represent the existing constraints on heavy non-annihilating DM interactions (see text for more details).}
		\label{fig:matched_vs_semi}
	\end{figure}
	\section{Acknowledgments} 
	We sincerely thank Basudeb Dasgupta, Ranjan Laha, and Nirmal Raj for helpful discussions and important inputs on the manuscript. S.B.  acknowledges ICGC-2023 conference at IIT Guwahati for its kind hospitality where part of this research was completed. A.L.M acknowledges the support for this research by the Netherlands Organisation for Scientific Research (NWO). A.R. acknowledges support from the National Science Foundation (Grant No. PHY-2020275) and to the Heising-Simons Foundation (Grant No. 2017-228).
	\bibliographystyle{JHEP}
	\bibliography{ref.bib,cw.bib}
\end{document}